\documentclass[twocolumn,showpacs,preprintnumbers,amsmath,amssymb,latexsym,aps]{revtex4}
\usepackage{graphicx}% Include figure files
\usepackage{dcolumn}% Align table columns on decimal point
\usepackage{bm}% bold math
\usepackage{color}

\definecolor{dblue}{rgb}{0,0,.75}
\begin{document}

%\preprint{APS/123-QED}

\title{Particles on curved surfaces - a dynamic approach by a phase field crystal model}

% Force line breaks with

\author{Rainer Backofen}
 \email{rainer.backofen@tu-dresden.de}
\author{Axel Voigt}
 \email{axel.voigt@tu-dresden.de}
\author{Thomas Witkowski}
 \email{thomas.witkowski@tu-dresden.de}
\affiliation{Department of Mathematics, Technische Universit\"at Dresden, 01062 Dresden, Germany}

\date{\today}% It is always \today, today,

\begin{abstract}
We present a dynamic model to study ordering of particles on arbitrary curved
surfaces. Thereby the particles are represented as maxima in a density field
and a surface partial differential equation for the density field is solved to
the minimal energy configuration. We study annihilation of 
dislocations within the ordered system and premelting along grain-boundary
scars. The obtained minimal energy configurations on a sphere are compared
with existing results and scaling laws are computed for the number of excess
dislocations as a function of system size. 
\end{abstract}

%\pacs{}% PACS, the Physics and Astronomy

\maketitle

Problems related to optimal ordering of particles on curved surfaces date back
to the classical Thomson problem \cite{Thomson_PM_1904} to find the ground
state of $N$ particles on a sphere interacting with a Coulomb potential. A
classic theorem of Euler shows for a triangulation of the surface in which
nearest neighbors are connected, that $\sum_i (6 - i) v_i = 6 \chi$, with
$v_i$ as the number of vertices with $i$ nearest neighbors and $\chi$ as the
Euler characteristic of the surface. Thus for surfaces with the topology of a
sphere ($\chi = 2$), besides the expected triangular lattice with six-fold
coordination, which would give the optimal packing in a plane, there must be
at least 12 five-fold disclinations present. With each disclination an extra
energy is associated (relative to a perfect triangular lattice in flat space)
which grows proportional to $r^2$, with $r$ as the radius of the sphere. For a
fixed lattice constant $a$ we have $N \sim (r/a)^2$. Thus for large $N$
mechanisms are expected which reduce this extra energy by changing the
ground-state configuration. One mechanism is a buckling transition of the
disclinations, which form sharp corners and turn the sphere into a
polygon. The transition depends on Young's modulus $Y$ and bending rigidity
$b$ via the Foppl-von Karman number $Y r^2 / b$ and is intensively studied for
viral capsids where protein subunits play the role of the particles
\cite{LindmarMirnyNelson_PRE_2003,ZandiRegueraBruismaGelbartRudnick_PNAS_2004}. In
cases where large surface tension limits significant buckling the energy can
be reduced by introducing grain-boundary scars. Realizations are for example
water droplets in oil, which are coated with colloidal particles
\cite{Bauschetal_Science_2003}. Such coated droplets are potential drug
delivery vehicles
\cite{Ramosetal_Science_1999,Dinsmoreetal_Science_2002}. Similar
configurations occur if a jammed layer of colloidal particles separats two
immiscible fluids forming a so-called bijel \cite{Stratfordetal_Science_2005},
which has potential applications as an efficient micro-reacting media.  A
large number of ordered particles on curved surfaces is also required for
fabrication of nanostructures on pliable substrates, e.g. to make foldable
electronic devices \cite{Sunetal_NatNano_2007}. For all these applications a
detailed understanding of the grain boundary scars is of interest as they may
be sources of leaks, influence mechanical properties or lead to failure in
electronic devices. We introduce an efficient way to compute the dynamics of
these grain boundary scars and dislocations associated with them and provide
an approach to compute optimal ordering of many particles on arbitrarily
curved surfaces. As the grain boundary scars belong to the thermal and
mechanical equilibrium our approach is based on energy minimization with the
geometric frustration resulting from the curved surface incorporated. 

For $2 \leq N \leq 100$ there is agreement of all numerical and theoretical
methods for the Thomson problem, suggesting that the global minimum
configuration has been found. However, for large $N$, owing to an exponential
growth in local minima \cite{ErberHockney_JPA_1991}, finding global minima
becomes extremely difficult. Grain boundary scars are expected for $N >
360$. Numerical approaches to solve such problems are typically based on
genetic algorithms, steepest decent minimization or coarse grained approaches,
in which the elasticity field between grain boundary scars is solved
\cite{BowickNelsonTravesset_PRB_2000,BowickCacciutoNelsonTravesset_PRL_2002}. All
approaches are devoted to finding the ground state. Dynamic models have been
considered \cite{Bowicketal_PRE_2007} which allow us to describe
experimentally observed dislocation glide within the grain boundary scars
\cite{Lipowskietal_NatMat_2005,Ling_NatMat_2005}. We will introduce an approach without any coarse graining by directly addressing the dynamic evolution and rearrangement of the particles on an arbitrary curved surface. Our approach is based on a free energy functional for a number density. In the plane such free energy functionals have been used to characterize patterns. The simplest possible form of a free energy which produces periodic structures in a domain $\Omega$ reads 
\begin{eqnarray}
\label{eq:free_energy}
{\cal F}[\rho] = \int_\Omega - |\nabla \rho|^2 + \frac{1}{2} |\Delta \rho|^2 + f(\rho) \; d\Omega
\end{eqnarray}
with $\rho$ as the number density and $f(\rho) = \frac{1}{2}(1 - \epsilon)\rho^2
+ \frac{1}{4} \rho^4$ as a potential with a parameter $\epsilon$. The
equilibrium state for $\Omega = {\mathbb R}^2$ has a perfect six-fold
symmetry. Evolutional laws associated with this energy are the $L^2$-gradient
flow $\partial_t \rho = - \delta {\cal F} / \delta \rho$, the Swift-Hohenberg
model \cite{SwiftHohenberg_PRA_1977}, and the $H^{-1}$ gradient flow
$\partial_t \rho = \Delta \delta {\cal F} / \delta \rho$, the phase field
crystal (PFC) model \cite{ElderKatakowskiHaatajaGrant_PRL_2002}. The
evolutions naturally contain elastic energy, as an expansion of the free
energy around the equilibrium period spacing results in the potential energy
of a spring, i\.e\. Hooke's law. As the energy is rotationally invariant
arbitrary orientationtions of periodic structures can emerge. Furthermore the
model allows the formation of dislocations, which occur when two periodic
structures of different orientation collide or when it is energetically
favorable for them to nucleate. The PFC model has been used to simulate
various crystal growth phenomena including epitaxial growth, nucleation,
commensurate-incommensurate transitions and plastic deformations. In
\cite{TeeffelenBackofenVoigtLoewen_PRE_2009} is shown how the model can be
derived from a microscopic Smoluchowski equation via dynamical density
functional theory. 
Formulating the energy in Eq. (\ref{eq:free_energy}) on a curved surface $\Gamma$ leads to 
\begin{eqnarray}
\label{eq:free_energy_surface}
{\cal F}^\Gamma[\rho] = \int_\Gamma - |\nabla_\Gamma \rho|^2 + \frac{1}{2} |\Delta_\Gamma \rho|^2 + f(\rho) \; d\Gamma
\end{eqnarray}
with $\rho$ as the number density on $\Gamma$, as well as $\nabla_\Gamma$ and
$\Delta_\Gamma$ as the surface gradient and surface Laplacian,
respectively. We will use the $H^{-1}$ gradient flow of this energy to solve
the generalized Thomson problem and to analyze the dynamics of rearrangements
of particles on a curved surface. The equation, written as a system of three
second order equations, reads as 
\begin{eqnarray}
\label{eq_1}
\partial_t \rho &=& \Delta_\Gamma u \\
\label{eq_2}
u &=&  \Delta_\Gamma v + 2\,v + f^\prime(\rho) \\
\label{eq_3}
v &=& \Delta_\Gamma \rho.
\end{eqnarray}
The stable finite element discretization for the PFC model in the plane
introduced in \cite{BackofenRaetzVoigt_PML_2007} can be adapted to solve
Eqs. (\ref{eq_1})-(\ref{eq_3}) on a surface triangulation using parametric
finite elements. The key idea is to use the surface operators on the discrete
surface which consists of triangles $T$. To do the integration on these
triangles a parametrization $F_{T}: \hat{T} \to T$ is used, with $\hat{T}$
as the standard element in $\mathbb{R}^2$. These allow us to transform all
integrations onto the standard element using the finite element basis defined
also only in $\mathbb R^{2}$. The parametrization $F_T$ is given by the
coordinates of the surface mesh elements and provides the only difference
between solving equations on surfaces and on planar domains. For a surface we
have to allow $F_T: \mathbb{R}^2 \to \mathbb{R}^{3}$, whereas for a planar
domain $F_T: \mathbb{R}^2 \to \mathbb{R}^2$. With this tiny modification any
code to solve partial differential 
equations on Cartesian grids can be used to solve the same problem on a
surface, providing a surface triangulation is given. The computational cost is
the same as solving the problem in a planar domain. Within an efficient
implementation this does not even require to recompile a running
two-dimensional (2D) code, but
only a change in a parameter file, as e.g. done in AMDiS
\cite{VeyVoigt_CVS_2007}. With this approach all available tools to solve
partial differential equations on planar domains, such as adaptive refinement,
multigrid algorithms or parallelization approaches are available also to solve
equations on surfaces. 
\begin{figure}[hbtp]
\begin{center}
\includegraphics[width=4cm]{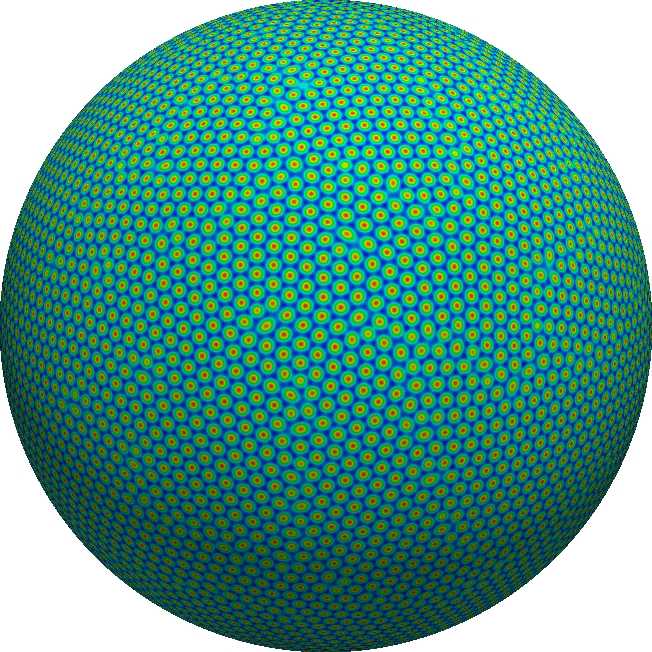}
\includegraphics[width=4cm]{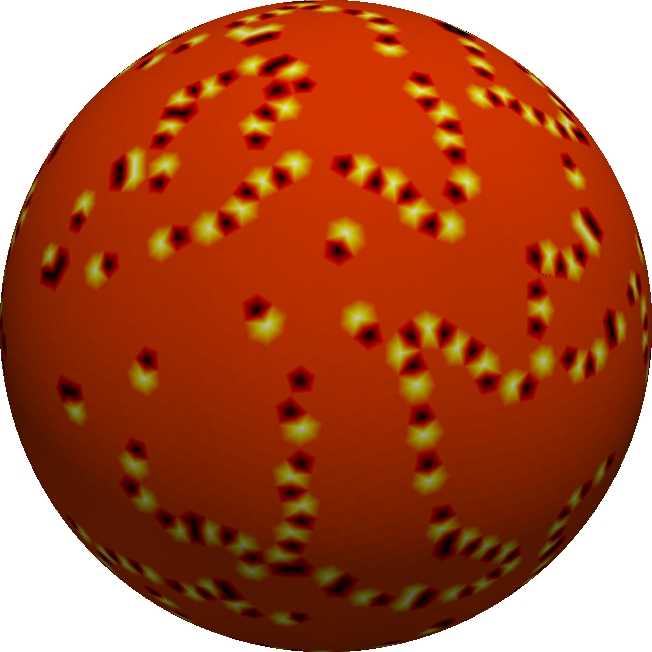}
\caption{(Color online) Local minimal energy configuration for 6.064 particles on a sphere. (left) density profile, (right) color coded number of neighbors, 5-black, 6-red (gray) and 7-yellow (white).} 
\label{fig:1}
\end{center}
\end{figure}

The approach is used to evolve a randomly perturbed constant initial
configuration $\rho = \rho_0$ towards an energy minimum. With the
possibilities to use adaptive time stepping and the efficiency of
parallelization of the finite element method problem sizes of $1 \times 10^6$
particles can be addressed. The simulation for Fig. \ref{fig:1} with 6.064
particles required 1 day computing time on a single processor.  
\begin{table}[b]
%  \begin{center}
   \caption{Comparison with known results for small $N$. In order to compute
     the energy we identify the position of the maxima in the denisty field
     and compute the Coulomb energy according to this positions.} 
    \begin{tabular}{rrrrrrr} \\
      \hline
      \textbf{N} & \textbf{v4} & \textbf{v5} & \textbf{v6} & \textbf{v7} &
      \textbf{v8} & \textbf{energy} \\ \hline 
      63 & 0 & 12 & 51 & 0 & 0 &         1708.87968150 \\ \hline
      99 & 0 & 12 & 87 & 0 & 0 &         4357.13916313 \\ \hline
      130 & 0 &	13 & 116 & 1 & 0 &       7632.16737891 \\ \hline
      185 & 0 &	12 & 173 & 0 & 0 &      15723.72346397 \\ \hline
      222 & 0 &	13 & 208 & 1 & 0 &      22816.07553076 \\ \hline
      363 & 0 &	14 & 347 & 2 & 0 &      62066.53633167 \\ \hline
      684 & 0 &	33 & 630 & 21 & 0 &    224048.60512144 \\ \hline
      846 & 0 &	38 & 782 & 26 &	0 &    344267.84965308 \\ \hline
      1073 & 0 & 45 & 995 & 33 & 0 &   556250.19927822 \\ \hline
      1403 & 0 & 54 & 1307 & 42 & 0 &  955173.65896550 \\ \hline
      2726 & 0 & 78 & 2584 & 64	& 0 & 3636897.41372145 \\ \hline
    \end{tabular}
     \label{tab:1}
%  \end{center}
\end{table}

In order to validate our approach we compute minimal energy configurations for
various numbers of $N$. We systematically compute the minimal energy
configuration for all $N \in [12, 2790]$. In our configuration $N = 2790$
corresponds to $r = 100$. The numerical results indicate, that the obtained
minimal energy configurations are only sensitive to the defined lattice
spacing and insensitive to a large extent to the other parameters. This might
explain why triangular tessellations on spherical surfaces occur in very
distinct occasions, for which the interactions involved may differ a lot. In
the following we use $\rho_0 = -0.3$ and $\epsilon = 0.4$, which together with
the radius $r$ of the sphere determines $N$. For all $N$ the type and number of
disclinations, as well as the computed energy is in agreement with known
analytical or other numerical results. For $N \leq 112$ the configurations and
energies coincide with the known equilibrium values. The maximal deviation
from the minimal energy for larger $N$ is less than 0.1 \%. Table \ref{tab:1}
shows computed configurations for selected $N$. 
\begin{figure}[t]
\begin{center}
\includegraphics[width=2.6cm]{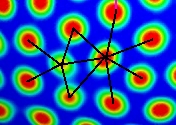}
\includegraphics[width=2.6cm]{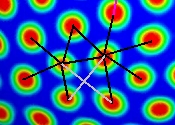}
\includegraphics[width=2.6cm]{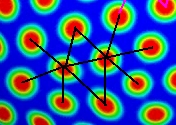}
\caption{(Color online) Annihilation of dislocation by local rearrangment of 5-7 defect. The gray lines indicate the formation of a new neighboure at an intermediate state.} \label{fig:3} 
\end{center}
\end{figure}
For $N > 360$ we obtain additional defects in the ground state, which are
pairs of fivefold and sevenfold coordinated particles (dislocations) and chains of
alternation fivefold and sevenfold coordinated particles (grain-boundary scars). Since
dislocations have vanishing total disclination charge there can be an
arbitrary number of them in any spherical lattice configuration without
violating the topological constraint on the total disclination charge
discussed above. The grain-boundary scars found are unlike any grain
boundaries found in flat space as they terminate freely inside the crystal at
both ends. Due to the importance of dislocations and grain boundaries in bulk
materials in determining material properties similar roles can be expected on
surfaces. In \cite{Lipowskietal_NatMat_2005} the motion of dislocations is
observed experimentally. Dislocation motion can be separated into glide and
climb, where glide is motion parallel to the dislocation's Burger vector and
requires only local rearrangement of the lattice, and climb is motion
perpendicular to the Burger's vector and requires the presence of vacancies
and interstitials. In agreement with the observations in
\cite{Lipowskietal_NatMat_2005} and the computational results based on
elasticity in \cite{Bowicketal_PRE_2007} we also observe only glide motion
which leads to significant shape changes of the scars. Fig. \ref{fig:3} shows a
local rearrangement of a dislocation. 

\begin{figure}[hbtp]
\begin{center}
\includegraphics[width=2.6cm]{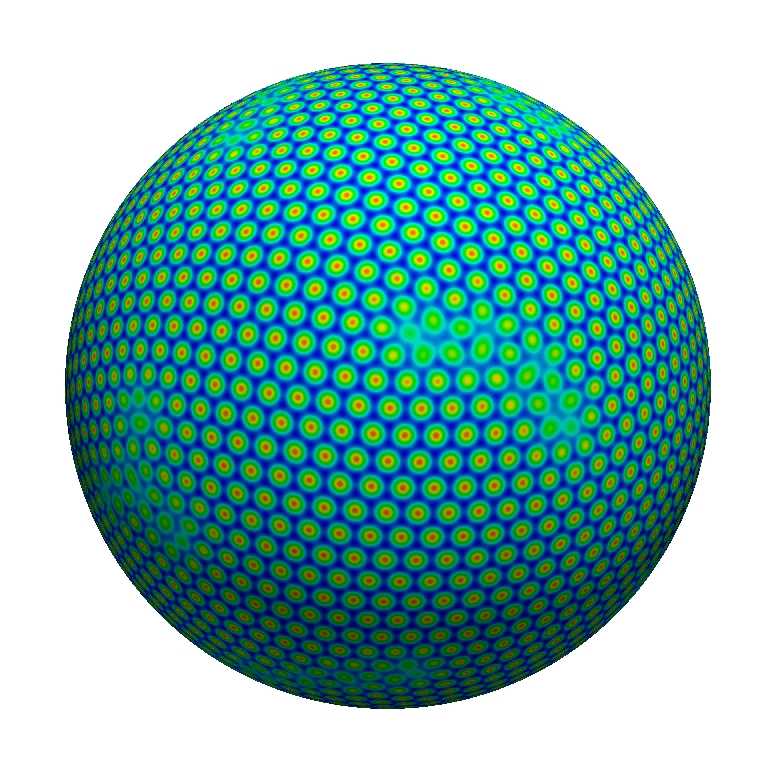}
\includegraphics[width=2.6cm]{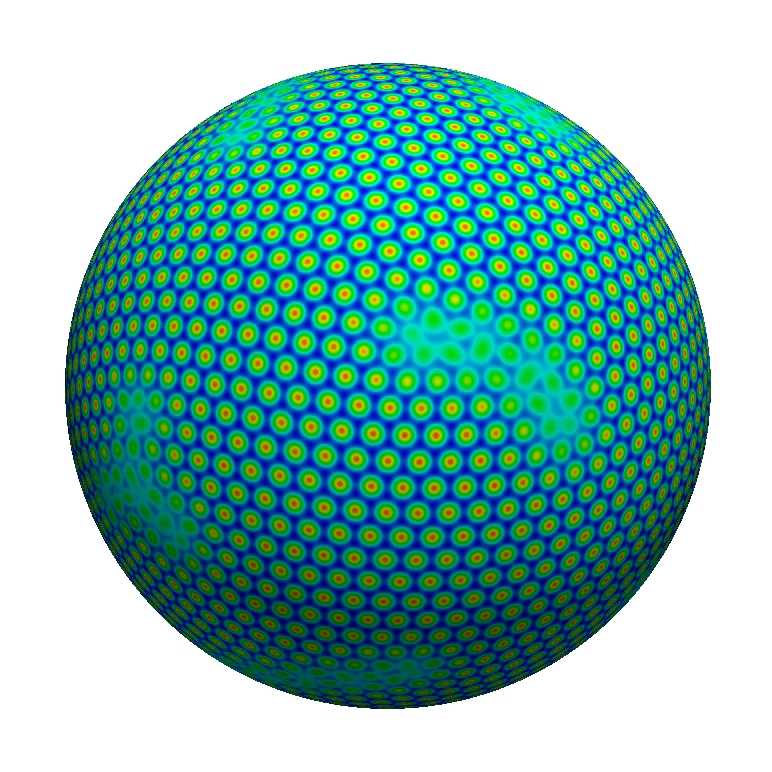}
\includegraphics[width=2.6cm]{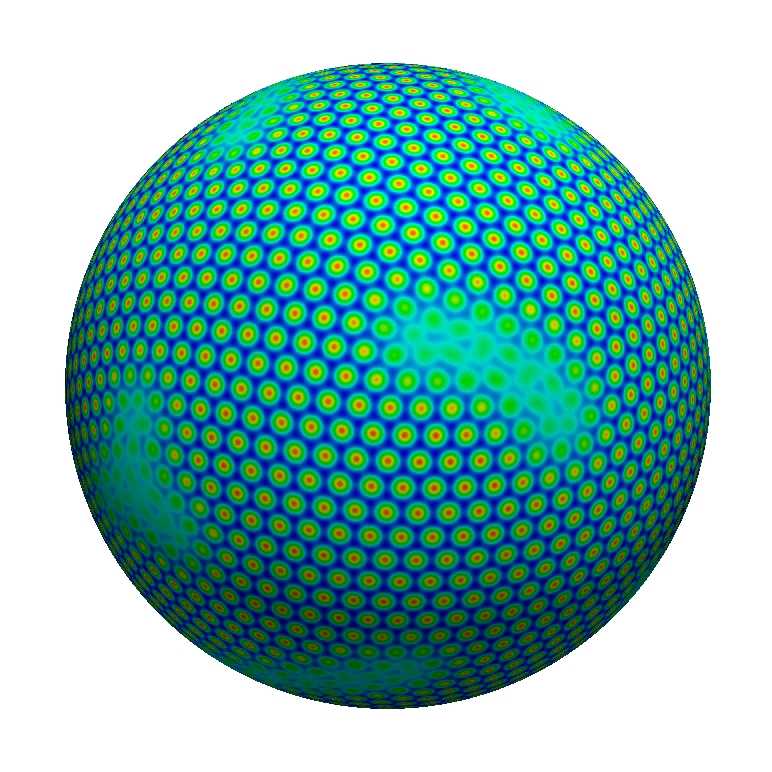}

\includegraphics[width=2.6cm]{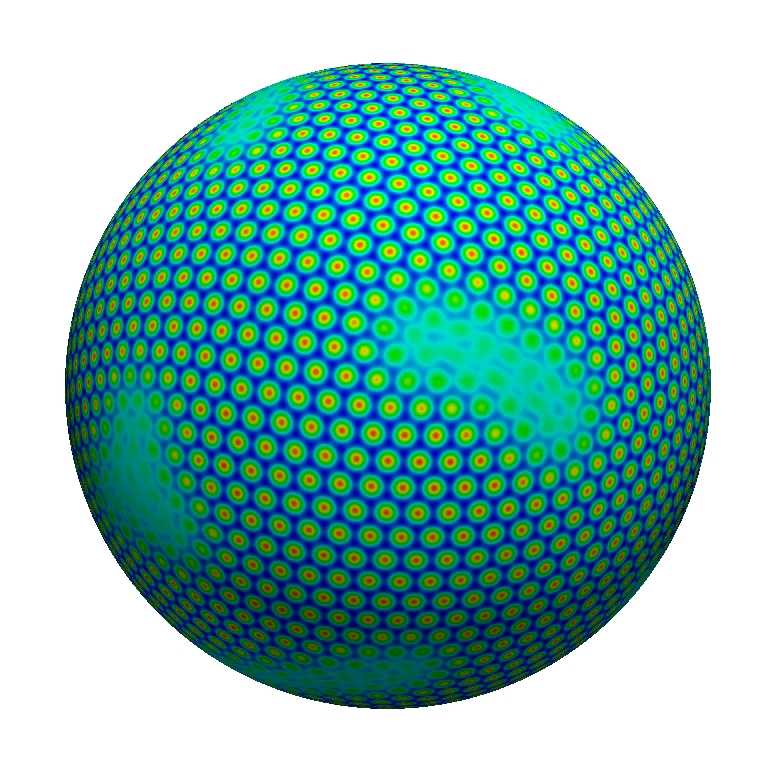}
\includegraphics[width=2.6cm]{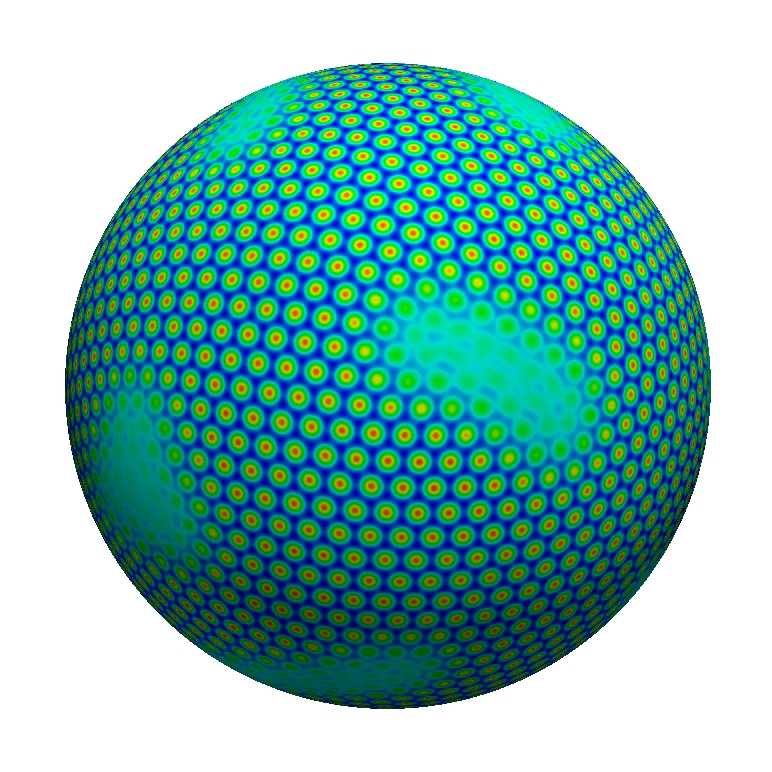}
\includegraphics[width=2.6cm]{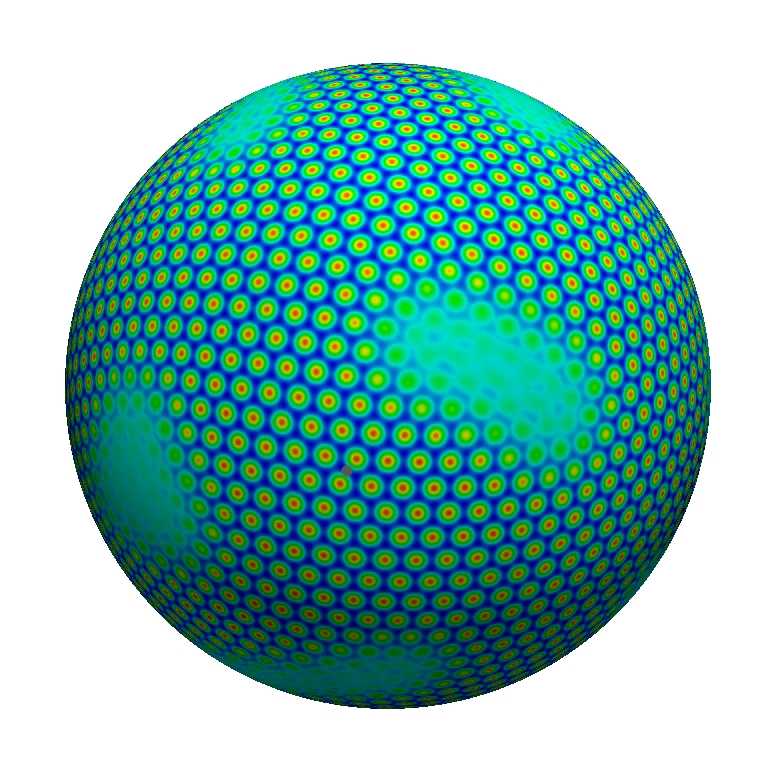}

\caption{(Color online) Time sequence showing premelting at grain boundary
  scars. The simulations indicate an initiation of the melting at the ends of
  the scars.}  
\label{fig:4} 
\end{center}
\end{figure}

In \cite{Ling_NatMat_2005} the question is asked which effect a raising of
temperature has on the spherical crystal. For bulk polycrystalline material
there is indirect experimental evidence for the occurrence of grain boundary
premelting, which could directly be visualized for colloidal crystals in
\cite{Alsayed_Science_2005}. As a liquid film at grain boundaries will alter
macroscopic properties and especially will lead to a drastically reduced
resistance to shear stresses which can lead to material failure it is not only
of theoretical interest if grain-boundary premelting is also present in
spherical crystals. As grain-boundary scars belong to the equilibrium state it
would be even more severe as it would be a general property of crystalline
materials on curved surfaces. PFC models have already been used to study 
grain-boundary premelting in flat space
\cite{BerryElderGrant_PRB_2008,MellenthinKarmaPlapp_PRB_2008}. For high-angle
grain boundaries an uniformly wetting is observed below the melting temperature
in these studies. Raising the temperature (increasing $\epsilon$) but keeping
it below the melting temperature according to the phase diagram indeed leads
to melting of the grain-boundary scars, see Fig. \ref{fig:4}.  

\begin{figure}[hbtp]
\begin{center}
\includegraphics[width=8cm]{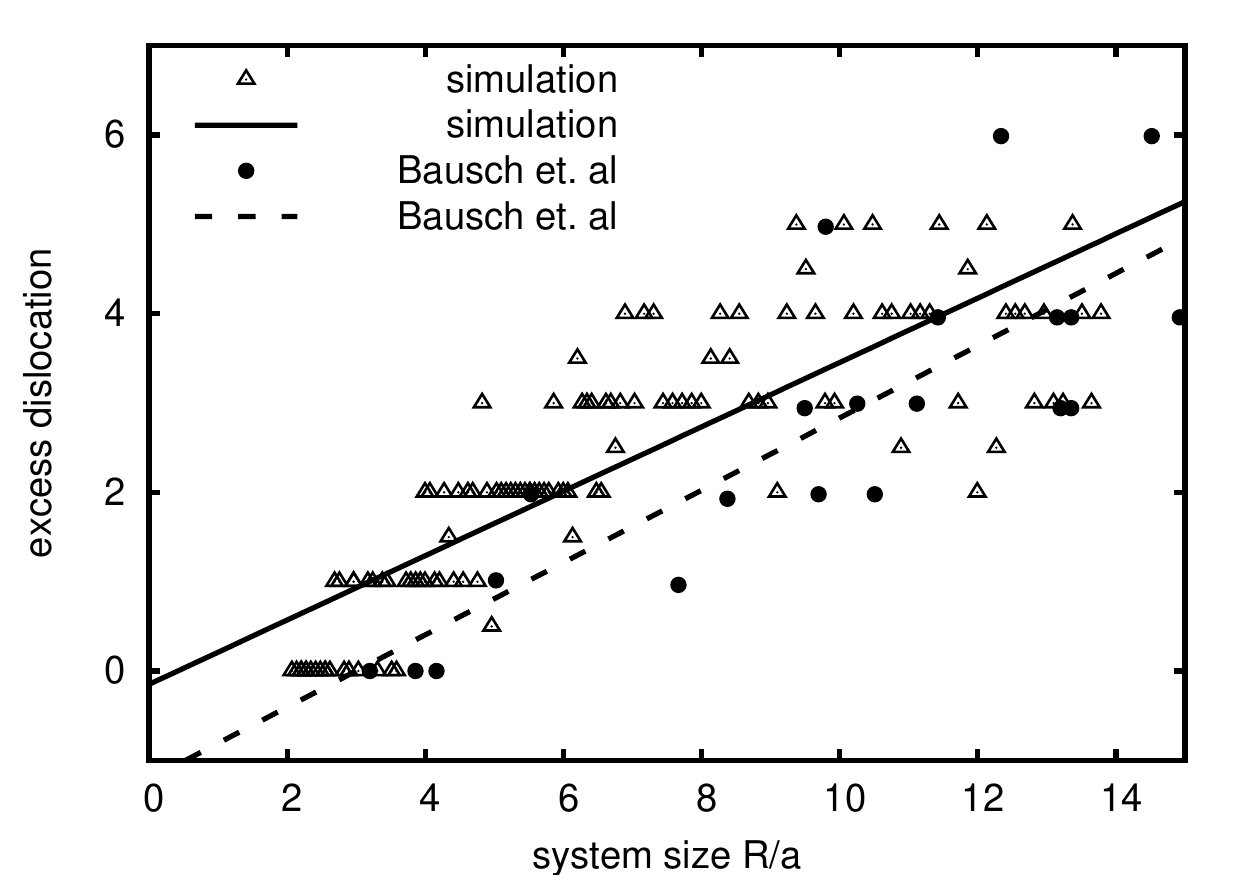}
\caption{Excess dislocations as a function of system size. The obtained slope
  is $0.388 \pm 0.020$, which is in excelent agreement with the experimental
  measured  
data in \cite{Bauschetal_Science_2003}, which give $0.404 \pm 0.062$, and the
theoretical value of $0.41$ from \cite{BowickNelsonTravesset_PRB_2000}}  
\label{fig:5} 
\end{center}
\end{figure}

We use the premelting of the dislocations to improve the local minima our
evolution settles in. Within an iterative procedure we evolve the system
according to Eqs. (\ref{eq_1}) - (\ref{eq_3}) until we reach a minimal state,
increase the temperature and run the system until liquid layers have replaced
all dislocations, 
and start the evolution again with the original temperature. The algorithm is
terminated if the total energy does not further decrease. Typically this is
achieved after 2-3 iteration cycles. In Figure \ref{fig:5} we plot the excess
dislocations in a scar as a function of the system size $\sqrt{N} = r / a$. 

%\begin{figure}[hbtp]
%\begin{center}
%\includegraphics[width=6.5cm]{fig6.png}
%\caption{Minimal energy configuration on Stanford bunny. Selected defects with 5- and 7-fold disclinations are %marked.} 
%\label{fig:6b} 
%\end{center}
%\end{figure}

\begin{figure}[hbtp]
\includegraphics[width=4.cm]{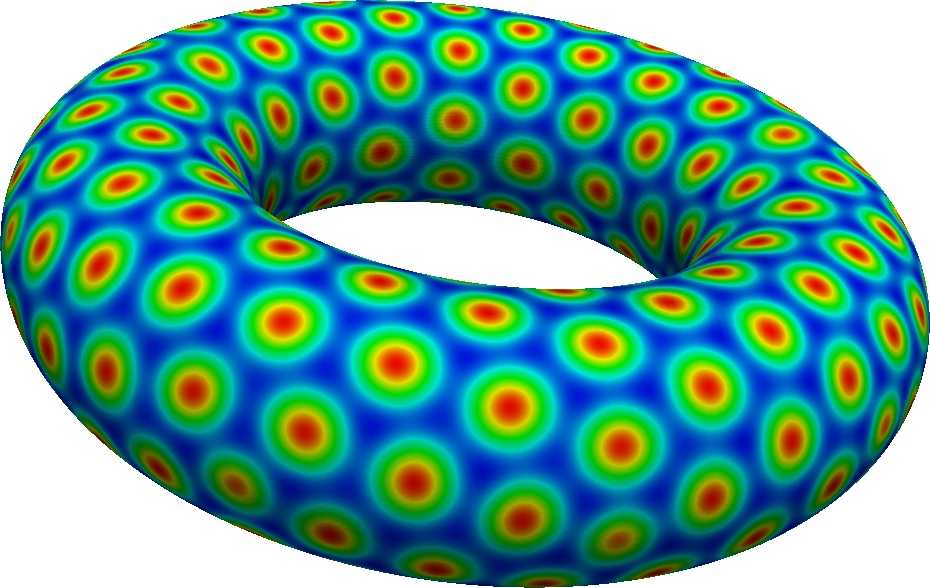}
\includegraphics[width=4.cm]{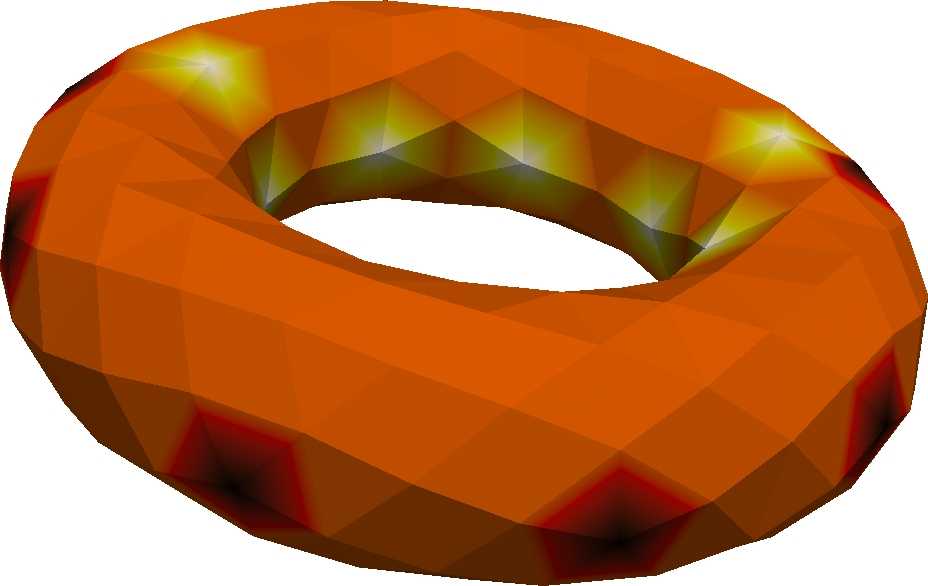}
\caption{(Color online) Minimal energy configuration on torus with $N = 239$,
  there are 13 fivefold disclinations and 13 sevenfold disclinations, 5-black,
  6-red (gray) and 7-yellow (white).} \label{fig:7}  
\end{figure}

As already pointed out the method is not restricted to spherical geometries. Indeed the algorithm works for arbitrary surfaces with the only requirement that an appropriate surface mesh is needed on which the computation can be done. As an example we use toroidal crystals, which can be found e.g. in self-assembled monolayers of micelles and vesicles \cite{Kimetal_JACS_2006} or carbon nanotori
\cite{Liuetal_Nature_1997}. We compute low energy configurations for various toroidal lattice configurations and reach good comparison with results reported in \cite{GiomiBower_PRE_2008}. Fig. \ref{fig:7} shows a typical configuration for aspect ratio $R_1/R_2 = 2.78$. 
 
In computations on more complicated surfaces, with convex and concave regions,
we observe isolated fivefold and sevenfold disclinations which arrange
according to the 
local curvature of the surface. Thereby fivefold disclinations are preferably
found in convex regions, whereas sevenfold disclinations are present in concave
regions, which is in accordance with the theory discussed in
\cite{Vitellietal_PNAS_2006}. The ability of the approach to work on arbitrary
surfaces will also allow us to consider ordering on evolving surfaces. As
discussed above for low surface tensions a buckling transition of the
disclinations can turn the sphere into a polygon inorder to reduce the
energy. The question arises if such a transition can intervene with 
grain boundary scar formation. In \cite{IoriSen_CEJB_2008} it is speculated
that grain boundary scars could be formed on capsids at an intermediate stage
of their evolution and that the release of the bending energy present in these
scars into stretching energy could allow for shape changes. To model such
shape changes requires us
to evolve the surface. Appropriate continuum models which account for bending
and surface tension are discussed in the mathematical review papers
\cite{DeckelnickDziukElliott_AN_2005,LiLowengrubRaetzVoigt_CCP_2009}. Different
concepts have been developed to solve differential equations on evolving
surfaces, 
\cite{DziukElliott_IMAJNA_2007} in the context of parametric finite elements
(as considered here) and \cite{RaetzVoigt_Nonlin_2007} within a phase-field
context. The coupling of the surface evolution with the evolution of the
number density on it and with it the question if grain boundary scars can
initiate a buckling transition, however remains open and requires further
developments. 

\begin{acknowledgments}
This work was supported by German Science Foundation Grants No. VO899/6-1 and
No. VO899/6-2.
\end{acknowledgments}

\end{document}